\documentclass[12pt]{article}
\usepackage{a4wide}
\usepackage{latexsym}
\usepackage{cite}

\usepackage{pslatex}
\usepackage[latin1]{inputenc}
\usepackage[T1]{fontenc}

\def\bq{\begin{eqnarray}}
\def\eq{\end{eqnarray}}

\def\eps{\varepsilon}

\begin{document}

\thispagestyle{empty}

\begin{flushright}
  MPP-2004-22
\end{flushright}

\vspace{1.5cm}

\begin{center}
  {\Large\bf Expansion around half-integer values, binomial sums and inverse binomial sums\\
  }
  \vspace{1cm}
  {\large Stefan Weinzierl\footnote{Heisenberg fellow of the Deutsche Forschungsgemeinschaft}\\
  \vspace{1cm}
  {\small {\em Max-Planck-Institut f\"ur Physik (Werner-Heisenberg-Institut),\\
               F\"ohringer Ring 6, D - 80805 M\"unchen, Germany}}
  } \\
\end{center}

\vspace{2cm}

\begin{abstract}\noindent
  {
I consider the expansion of transcendental functions in a small parameter around
rational numbers.
This includes in particular the expansion around half-integer values.
I present algorithms which are suitable for an implementation within a symbolic computer
algebra system.
The method is an extension of the technique of nested sums.
The algorithms allow in addition the evaluation of binomial sums, inverse binomial sums
and generalizations thereof.
   }
\end{abstract}

\vspace*{\fill}

\newpage

\section{Introduction}
\label{sec:intro}

The expansion of higher transcendental functions \cite{Erdelyi,Slater}
occurs frequently in many areas of science.
In particular, one encounters these functions in the calculation of higher order
corrections to scattering processes in particle physics.
In a previous publication we considered the expansion of transcendental functions
in a small parameter around integer values
\cite{Moch:2001zr,Weinzierl:2002hv}.
The restriction to integer values is in general sufficient for the evaluation
of loop integrals arising in massless quantum field theories.
However, the inclusion of particle masses in loop integrals 
\cite{Fleischer:1998nb,Jegerlehner:2002em,Davydychev:2000na,Kalmykov:2000qe,Davydychev:2003mv,Fleischer:1999mp,Davydychev:1999mq,Bonciani:2003hc,Aglietti:2004tq}
or the evaluation
of phase space integrals 
\cite{Phaf:2001gc,Kosower:2002su,Kosower:2003cz,Weinzierl:2003fx,Weinzierl:2003ra,Gehrmann-DeRidder:2003bm,Anastasiou:2003gr}
can lead to half-integer values.
It is therefore desirable to extend the algorithm of \cite{Moch:2001zr,Weinzierl:2002hv}
to include at least half-integer values.
Here I report on algorithms for the expansion of transcendental functions around rational
number $p/q$, where $p$ and $q$ are integers.
In particular this includes the half-integer case.

Each term in the expansion is expressed through multiple polylogarithms
\cite{Goncharov,Borwein}.
Compared to the pure integer case, the extension to rational numbers $p/q$
introduces naturally the $q$-th roots of unity in the arguments of the polylogarithms.
All algorithms are based on manipulations of a special form of nested sums.
These nested sums are generalizations of Euler-Zagier sums 
\cite{Euler,Zagier} 
or harmonic sums
\cite{Gonzalez-Arroyo:1979df,Gonzalez-Arroyo:1980he,Vermaseren:1998uu,Blumlein:1998if,Blumlein:2003gb}.
The algorithms presented here can be implemented into a symbolic computer algebra system
like Form \cite{Vermaseren:2000nd,Vermaseren:1998uu}
or GiNaC \cite{Bauer:2000cp}.

As a spin-off, the methods presented here allow the evaluation of binomial sums
\cite{Fleischer:1998nb,Jegerlehner:2002em}, 
inverse binomial sums 
\cite{Ogreid:1998bx,Fleischer:1998nb,Borwein:2000et,Davydychev:2000na,Kalmykov:2000qe,Davydychev:2003mv}
and generalizations thereof.
Inverse binomial sums are sometimes evaluated with the help of log-sine integrals
\cite{Fleischer:1999mp,Davydychev:1999mq}.
In an appendix I compare the log-sine approach with the one presented here.

This paper is organized as follows:
Sect. \ref{sec:review} recalls the definition and main properties of nested sums
and multiple polylogarithms. It is a brief summary of ref. \cite{Moch:2001zr}.
Sect. \ref{sec:algos} introduces roots of unity and gives the basic algorithms for the expansion
around rational numbers.
Sect. \ref{sec:binomial} treats binomial sums and generalizations thereof.
Sect. \ref{sec:invbinomial} deals with inverse binomial sums and generalizations thereof.
Sect. \ref{sec:appl} gives some simple applications to massive loop integrals and phase space integrals.
Finally, sect. \ref{sec:concl} contains a summary and the conclusions.
An appendix compares this approach to log-sine integrals and collects some important
relations for polylogarithms of low weight.


\section{A summary of known properties of nested sums}
\label{sec:review}

In this section I shortly review properties of particular forms of
nested sums, which are called $Z$-sums and $S$-sums.
Details can be found in \cite{Moch:2001zr}.
$Z$-sums are defined by
\bq 
  Z(n;m_1,...,m_k;x_1,...,x_k) & = & \sum\limits_{n\ge i_1>i_2>\ldots>i_k>0}
     \frac{x_1^{i_1}}{{i_1}^{m_1}}\ldots \frac{x_k^{i_k}}{{i_k}^{m_k}}
\eq
and form a Hopf algebra.
If the sums go to Infinity ($n=\infty$) the $Z$-sums are identical 
to multiple polylogarithms\footnote{Note that we use here the reversed notation
for multiple polylogs and multiple zeta values as compared to refs. \cite{Goncharov,Moch:2001zr}.}
\cite{Goncharov}:
\bq
\label{multipolylog}
Z(\infty;m_1,...,m_k;x_1,...,x_k) & = & \mbox{Li}_{m_1,...,m_k}(x_1,...,x_k).
\eq
For $x_1=...=x_k=1$ the definition reduces to the Euler-Zagier sums \cite{Euler,Zagier}:
\bq
Z(n;m_1,...,m_k;1,...,1) & = & Z_{m_1,...,m_k}(n).
\eq
For $n=\infty$ and $x_1=...=x_k=1$ the sum is a multiple $\zeta$-value \cite{Borwein}:
\bq
Z(\infty;m_1,...,m_k;1,...,1) & = & \zeta_{m_1,...,m_k}.
\eq
The multiple polylogarithms contain as the notation already suggests as subsets 
the classical polylogarithms 
$
\mbox{Li}_n(x)
$ 
\cite{lewin:book},
as well as
Nielsen's generalized polylogarithms \cite{Nielsen}
\bq
S_{n,p}(x) & = & \mbox{Li}_{n+1,1,...,1}(x,\underbrace{1,...,1}_{p-1}),
\eq
the harmonic polylogarithms \cite{Remiddi:1999ew}
\bq
\label{harmpolylog}
H_{m_1,...,m_k}(x) & = & \mbox{Li}_{m_1,...,m_k}(x,\underbrace{1,...,1}_{k-1})
\eq
and two-dimensional harmonic polylogarithms
\cite{Gehrmann:2000zt}.
The usefulness of the $Z$-sums lies in the fact, that they interpolate between
multiple polylogarithms and Euler-Zagier sums.
In addition, the interpolation is compatible with the algebra structure.

In addition to $Z$-sums, it is sometimes useful to introduce as well $S$-sums.
$S$-sums are defined by
\bq
S(n;m_1,...,m_k;x_1,...,x_k)  & = & 
\sum\limits_{n\ge i_1 \ge i_2\ge \ldots\ge i_k \ge 1}
\frac{x_1^{i_1}}{{i_1}^{m_1}}\ldots \frac{x_k^{i_k}}{{i_k}^{m_k}}
\eq
and form an algebra.
The $S$-sums reduce for $x_1=...=x_k=1$ (and positive $m_i$) to harmonic sums \cite{Vermaseren:1998uu}:
\bq
S(n;m_1,...,m_k;1,...,1) & = & S_{m_1,...,m_k}(n).
\eq
The $S$-sums are closely related to the $Z$-sums, the difference being the upper summation boundary
for the nested sums: $(i-1)$ for $Z$-sums, $i$ for $S$-sums.
It is advantageous to introduce both $Z$-sums and $S$-sums, since some 
properties are more naturally expressed in terms of
$Z$-sums while others are more naturally expressed in terms of $S$-sums.
One can easily convert from one notation to the other.
\\
\\
Basic manipulations involving nested sums are:
\\
Conversion:
\bq
\label{manipconv}
 Z(n;...) & \rightarrow & S(n;...), \nonumber \\
 S(n;...) & \rightarrow & Z(n;...).
\eq
Multiplication:
\bq
\label{manipmult}
 Z(n;...) \; Z(n;...) & \rightarrow & Z(n;...), \nonumber \\
 S(n;...) \; S(n;...) & \rightarrow & S(n;...).
\eq
Convolution:
\bq
\label{manipconvolut}
 \sum\limits_{i=1}^{n-1} \; \frac{x^i}{i^m} Z(i-1;...)
                         \; \frac{y^{n-i}}{(n-i)^{m'}} Z(n-i-1;...)
 & \rightarrow & Z(n-1;...).
\eq
Conjugation:
\bq
\label{manipconj}
 - \sum\limits_{i=1}^n 
       \left( \begin{array}{c} n \\ i \\ \end{array} \right)
       \left( -1 \right)^i
       \; \frac{x^i}{i^m} S(i;...)
 & \rightarrow & S(n;...).
\eq
Conjugation and convolution:
\bq
\label{manipconvconj}
 - \sum\limits_{i=1}^{n-1} 
       \left( \begin{array}{c} n \\ i \\ \end{array} \right)
       \left( -1 \right)^i
       \; \frac{x^i}{i^m} S(i;...)
       \; \frac{y^{n-i}}{(n-i)^{m'}} S(n-i;...)
 & \rightarrow & S(n-1;...).
\eq
These algorithms are described in
\cite{Moch:2001zr}.
It is worth to recall some technical steps to evaluate the conjugation in eq. (\ref{manipconj}), since the same
pattern of steps will be used in sect. \ref{sec:binomial}.
To evaluate the conjugation
it is convenient to introduce yet another type of sum as follows:
\bq
B(n;N;m_1,...,m_k;x_1,...,x_k) & = & 
 \sum\limits_{i_1=n+1}^N \sum\limits_{i_2=i_1+1}^N ... \sum\limits_{i_k=i_{k-1}+1}^N
   \frac{x_1^{i_1}}{i_1^{m_1}} \frac{x_2^{i_2}}{i_2^{m_2}} ... \frac{x_k^{i_k}}{i_k^{m_k}}.
\eq
These $B$-sums can be used to express $S$-sums with upper summation limit $n$ in terms of $S$-sums
with upper summation limit $N$:
\bq
\label{convertS2B}
\lefteqn{
    S(n;m_1,...,m_k;x_1,...,x_k) 
    = S(N;m_1,...,m_k;x_1,...,x_k) 
      - S(N;m_2,...,m_k;x_2,...,x_k) B(n;N;m_1;x_1) 
} & & \nonumber \\
& &
     + S(N;m_3,...,m_k;x_3,...,x_k) B(n;N;m_1,m_2;x_1,x_2)
     - ... + (-1)^k B(n;N;m_1,...,m_k;x_1,...,x_k).
 \nonumber \\
\eq
Eq. (\ref{convertS2B}) allows also
to express a $B$-sum recursively in terms of $S$-sums $S(N;...)$ and $S(n;...)$:
\bq
\lefteqn{
      B(n;N;m_1,...,m_k;x_1,...,x_k) = 
    (-1)^k S(n;m_1,...m_k;x_1,...,x_k) 
} & & \nonumber \\
& &
    - (-1)^k S(N;m_1,...m_k;x_1,...,x_k)
           + (-1)^k S(N;m_2,...m_k;x_2,...,x_k) B(n;N;m_1;x_1)
\nonumber \\
& &
           - ...
           + (-1)^k S(N;m_k;x_k) B(n;N;m_1,...,m_{k-1};x_1,...,x_{k-1})
\eq
Finally, it is convenient to
introduce raising and lowering operators as follows:
\bq
\left( {\bf x^+} \right)^m \cdot 1 & = & \frac{1}{m!} \ln^m(x), \nonumber \\
{\bf x^+} \cdot f(x) & = & \int\limits_0^x \frac{dx'}{x'} f(x'), \nonumber \\ 
{\bf x^-} \cdot f(x) & = & x \frac{d}{dx} f(x). 
\eq
It is understood that in the second line only functions which are integrable at $x=0$ are
considered. 
With the help of raising operators a $B$-sum $B(n;\infty;...)$ may be expressed as follows:
\bq
\lefteqn{
B(n;\infty;m_1,...,m_k;x_1,...,x_k) = }
\nonumber \\
 & &
     \left( {\bf x_k}^+ \right)^{m_k} \left( {\bf x_{k-1}}^+ \right)^{m_{k-1}} ... 
     \left( {\bf x_1}^+ \right)^{m_1} 
      \frac{x_k}{1-x_k} \frac{x_{k-1}x_k}{1-x_{k-1}x_k} ...
     \frac{x_1 ... x_k}{1-x_1 ... x_k}
     \left( x_1 ... x_k \right)^n.
\eq
Some important integrals related to the raising operators are:
\bq
\label{raisint}
\lefteqn{
{\bf x_1^+} \left[ 1 - \left( 1 - x_1 x_2 \right)^{n} \right] 
 = \sum\limits_{i=1}^n \frac{1}{i} \left[ 1 - \left( 1- x_1 x_2 \right)^i \right], 
} & & \nonumber \\
\lefteqn{
{\bf x_1^+} \frac{x_1 x_2}{1-x_1 x_2} \left[ 1 - \left( 1 - x_0 x_1 x_2 \right)^n \right] 
 =  
   - \left( 1 - x_0 \right)^n \sum\limits_{i=1}^n \frac{1}{i} 
                \left( \frac{1}{1-x_0} \right)^i \left[ 1 - \left( 1 - x_0 x_1 x_2 \right)^i \right] 
} & & \nonumber \\
& & 
    + \left( 1 - \left( 1 - x_0 \right)^n \right) \sum\limits_{i=1}^N \frac{\left(x_1 x_2 \right)^i}{i} 
+ {\bf x_1^+} \frac{x_1 x_2}{1-x_1 x_2} \left( x_1 x_2 \right)^N 
                                      \left( 1 - \left( 1 - x_0 \right)^n \right).
\eq
To evaluate the conjugation in eq. (\ref{manipconj}) one first converts the $S$-sum to $B$-sums and introduces
then the raising operators. This allows to perform all sums explicitly and one is left with the integrals
corresponding to the raising operators.
With the help of eq. (\ref{raisint}) one can systematically perform these integrals and convert them back
into nested sums.

The basic manipulations in eqs. (\ref{manipconv}) to (\ref{manipconvconj}) are the building blocks to reduce the following
two generic types of sums to single $Z$-sums.
\\
Type A:
\bq
\label{algo_A}
     \sum\limits_{i=1}^n \frac{x^i}{(i+c)^m} 
       \frac{\Gamma(i+a_1+b_1\varepsilon)}{\Gamma(i+c_1+d_1\varepsilon)} ...
       \frac{\Gamma(i+a_k+b_k\varepsilon)}{\Gamma(i+c_k+d_k\varepsilon)}
       Z(i+o-1,m_1,...,m_l,x_1,...,x_l)
\eq
Type B:
\bq
\label{algo_B}
\lefteqn{
     \sum\limits_{i=1}^{n-1} 
       \frac{x^i}{(i+c)^m} 
       \frac{\Gamma(i+a_1+b_1\varepsilon)}{\Gamma(i+c_1+d_1\varepsilon)}
...
       \frac{\Gamma(i+a_k+b_k\varepsilon)}{\Gamma(i+c_k+d_k\varepsilon)} \, 
       Z(i+o-1,m_1,...,m_l,x_1,...,x_l) }
\nonumber \\
& &
       \times
       \frac{y^{n-i}}{(n-i+c')^{m'}} 
       \frac{\Gamma(n-i+a_1'+b_1'\varepsilon)}{\Gamma(n-i+c_1'+d_1'\varepsilon)}
...
       \frac{\Gamma(n-i+a_{k'}'+b_{k'}'\varepsilon)}{\Gamma(n-i+c_{k'}'+d_{k'}'\varepsilon)}
\hspace*{4.0cm}
\nonumber \\[1ex]
& & \qquad\qquad \times
       Z(n-i+o'-1,m_1',...,m_{l'}',x_1',...,x_{l'}')
\eq
Here, all $a_j$, $a_j'$, $c_j$  and $c_j'$ are integers, $c$, $c'$, 
are nonnegative integers and $o $,
$o' $ are integers. For sums of type A the upper summation limit $n$ may extend to Infinity.
In the rest of the paper I relax the condition on $a_j$, $a_j'$, $c_j$  and $c_j'$ to allow for
rational numbers.
Ref. \cite{Moch:2001zr} contains in addition two generic types (labelled type C and D in \cite{Moch:2001zr}) 
involving a conjugation.
These types do not allow an extension to rational numbers along the lines of this paper.
The conjugation is however related to two other important types of sums, which are
generalizations of binomial and inverse binomial sums.
They are treated in sect. \ref{sec:binomial} and sect. \ref{sec:invbinomial} of this paper.


\section{Algorithms for the expansion around rational numbers}
\label{sec:algos}

In this section I extend the algorithms for the expansion of transcendental functions
towards the expansion around rational numbers.
Subsection \ref{subsect:rootofunity} introduces roots of unity.
Roots of unity are useful for the refinement algorithm in subsection \ref{subsect:refine},
which allows to express a $S$-sum $S(n;...)$ as a combination of $S$-sums, whose upper 
summation limit is an integer multiple of $n$.
Subsection \ref{subsect:gamma} treats the expansion of Euler's Gamma function around rational numbers.
Finally, in subsection \ref{subsect:expratnumb} all pieces are assembled and the algorithms for the expansion
of functions of type A and B around rational numbers are given.
A restriction on these algorithms is given by the fact, that rational numbers have to appear in the same 
place in the numerator and in the denominator.
This restriction is relaxed in sect. \ref{sec:binomial} (a rational number only in the numerator)
and sect. \ref{sec:invbinomial} (a rational number only in the denominator).

\subsection{Roots of unity}
\label{subsect:rootofunity}

We define a short-hand notation for the roots of unity:
\bq
r_q^p & = & \exp \left( \frac{2 \pi i p}{q} \right).
\eq
Here, $q$ is a positive integer and $p$ is a non-negative integer.
We will need a few properties of the $q$-th roots of unity.
Powers of the $q$-th roots of unity are periodic modulo $q$:
\bq
\left( r_q^p \right)^{j+q} & = & \left( r_q^p \right)^{j}.
\eq
Sums of powers of the $q$-th roots of unity yield:
\bq
\label{eq_roots_1}
\sum\limits_{p=0}^{q-1} \left( r_q^p \right)^{m} & = & 
 \left\{
  \begin{array}{cc}
    q, &  m = 0 \; \mbox{mod} \; q, \\
    0, &  m \neq 0 \; \mbox{mod} \; q.\\
  \end{array}
 \right.
\eq
If $m = 0\; \mbox{mod} \; q$, the proof of the relation is trivial.
In the case $m \neq 0\; \mbox{mod} \; q$ we may assume 
that $0 < m < q$ (due to the periodicity). Then $r_q^m \neq 1$ and we have
\bq
\sum\limits_{p=0}^{q-1} \left( r_q^p \right)^{m} 
 =
\sum\limits_{p=0}^{q-1} \left( r_q^m \right)^{p} 
 = 
 \frac{1- \left( r_q^m \right)^{q}}{1- r_q^m }
 = 0.
\eq
From eq. (\ref{eq_roots_1}) we obtain immediately
\bq
\label{eq_roots_2}
\frac{1}{q} \sum\limits_{l=0}^{q-1} \left( r_q^l \right)^{m+p} & = &
 \left\{
  \begin{array}{cl}
    1, &  \mbox{for} \;\; m = nq-p, \\
    0, &  \mbox{otherwise.}\\
  \end{array}
 \right.
\eq
Since this sum occurs frequently, we introduce the notation
\bq
\delta_{p,q}(m) & = &
\frac{1}{q} \sum\limits_{l=0}^{q-1} \left( r_q^l \right)^{m+p}.
\eq
Eq. (\ref{eq_roots_2}) will be useful to convert sums with upper summation limit $n$
to sums with upper summation limit $(q \cdot n)$.
Note that $\delta_{p,q}(m)$ is idempotent:
\bq
\delta_{p,q}(m) \cdot \delta_{p,q}(m) & = & \delta_{p,q}(m)\;\;\;\mbox{for all integer $m$.}
\eq
Some examples for the half-integer case are:
\bq
\delta_{0,2}(n) & = & \frac{1}{2} \left[ 1 + \left(-1\right)^n \right],
 \nonumber \\
\delta_{1,2}(n) & = & \frac{1}{2} \left[ 1 - \left( -1 \right)^n \right].
\eq

\subsection{Refinements of $S$-sums}
\label{subsect:refine}

A $S$-sum $S(n;m_1,...;x_1,...)$ 
with upper summation limit $n$ can be expressed as a combination
of $S$-sums with upper summation limit $(q \cdot n)$, where $q$ is a positive
integer.
The algorithm proceeds recursively in the depth of the $S$-sum. 
For the empty sum one has
\bq
S(n) & = & S( q \cdot n).
\eq
For a $S$-sum of the form
\bq
S(n;m_1,m_2,...;x_1,x_2,...) & = & 
 \sum\limits_{i=1}^n \frac{x_1^i}{i^{m_1}} S(i;m_2,...;x_2,...)
\eq
the algorithm converts first the subsum $S(i;m_2,...;x_2,...)$
to a combination of subsums $S(q \cdot i;...)$.
Finally, the outermost sum is converted according to
\bq
    \sum\limits_{i=1}^n \frac{x^i}{i^m} S( q \cdot i;...)
      = q^m \sum\limits_{i=1}^{q \cdot n} \delta_{0,q}(i) \frac{1}{i^m}
                \left( x^{1/q} \right)^i S(i;...)
      = q^{m-1} \sum\limits_{p=0}^{q-1}
                \sum\limits_{i=1}^{q \cdot n} \frac{1}{i^m}
                \left( r_q^p x^{1/q} \right)^i S(i;...).
\eq
This completes the algorithm for the conversion of $S$-sums $S(n;...)$
to $S(q \cdot n;...)$.
As an example we have
\bq
S(n;1,1;x_1,x_2) & = & 
 S( 2 n;1,1;\sqrt{x_1},\sqrt{x_2}) 
 + S( 2 n;1,1;\sqrt{x_1},-\sqrt{x_2}) 
\nonumber \\
& &
 + S( 2 n;1,1;-\sqrt{x_1},\sqrt{x_2}) 
 + S( 2 n;1,1;-\sqrt{x_1},-\sqrt{x_2}).
\eq
Note that as a consequence of the refinement algorithm one obtains
relations like
\bq
\mbox{Li}_m(x^2) & = & 2^{m-1} 
 \left[ \mbox{Li}_m(x) + \mbox{Li}_m(-x)\right].
\eq

\subsection{Expansion of the Gamma function}
\label{subsect:gamma}

For the expansion of the Gamma function around positive
integer values one has the well-known formula
\bq
\label{Gamma_integer}
\frac{\Gamma(n+1+\eps)}{\Gamma(1+\eps)} & = & 
\Gamma(n+1) \exp \left( - \sum\limits_{k=1}^\infty \eps^k \frac{(-1)^k}{k} S_k(n) \right).
\eq
For the expansion around rational numbers one finds
\bq
\label{Gamma_rational}
\frac{\Gamma\left( n+1-\frac{p}{q}+\eps \right)}
     {\Gamma\left( 1-\frac{p}{q}+\eps\right)}
 & = &
\frac{\Gamma\left( n+1-\frac{p}{q}\right)}{\Gamma\left( 1-\frac{p}{q} \right)}
\exp \left( - \sum\limits_{k=1}^\infty
             \eps^k \frac{(-q)^k}{k}
             \sum\limits_{j=1}^{q n} \frac{\delta_{p,q}(j)}{j^k}
      \right)
 \nonumber \\ 
 & = &
\frac{\Gamma\left( n+1-\frac{p}{q}\right)}{\Gamma\left( 1-\frac{p}{q} \right)}
\exp \left( - \frac{1}{q} \sum\limits_{l=0}^{q-1}
            \left( r_q^l \right)^p
            \sum\limits_{k=1}^\infty
             \eps^k \frac{(-q)^k}{k}
             S( q \cdot n; k; r_q^l )
      \right).
\eq
Here, $n$ and $q$ are positive integers and
$p$ is an integer with $0 \le p < q$ and $\mbox{gcd}(p,q) = 1$.
For the case $p=0$ and $q=1$ this reduces to the formula eq. (\ref{Gamma_integer}).
Eq. (\ref{Gamma_rational}) is derived as follows:
One starts from the expansion of the logarithm of the Gamma function:
\bq
\ln \Gamma\left( n+1-\frac{p}{q}+\eps \right) & = &
 \ln \Gamma\left( n+1-\frac{p}{q} \right)
 + \sum\limits_{k=1}^\infty \frac{\eps^k}{k!} 
                            \psi^{(k-1)}\left( n+1-\frac{p}{q} \right).
\eq
Here $\psi^{(k-1)}(x)$ is the polygamma function defined as
\bq
\psi^{(k-1)}(x) & = & \frac{d^k}{dx^k} \ln \Gamma(x).
\eq
From the recurrence formula for the polygamma function
\bq
\psi^{(k-1)}(x+1) -\psi^{(k-1)}(x)
 & = & - \Gamma(k) \left( - \frac{1}{x} \right)^k
\eq
one obtains
\bq
\label{eq_polygamma}
\psi^{(k-1)}\left( n+1-\frac{p}{q} \right) & = &
  \psi^{(k-1)}\left( 1-\frac{p}{q} \right)
  - \Gamma(k) (-1)^k \sum\limits_{i=1}^n 
                     \left( \frac{1}{i - \frac{p}{q}} \right)^{k}.
\eq
The sum in the last term of eq. (\ref{eq_polygamma}) is then rewritten in terms
of $S$-sums with upper summation limit $(q \cdot n)$ with the help
of eq. (\ref{eq_roots_2}).

\subsection{Expansion around rational numbers}
\label{subsect:expratnumb}

In this section we generalize the algorithms A and B, which expand
the transcendental sums in eq. (\ref{algo_A}) and eq. (\ref{algo_B})
around integers to algorithms for the expansion around rational
numbers.
In the following we will always assume that $q_j$ is a positive integer
and that $p_j$ is an integer with $0 \le p_j < q_j$ and
$\mbox{gcd}(p_j,q_j) = 1$.
We restrict ourselves here to ratios of Gamma function of the form
\bq
\label{restrictionnumdenom}
\frac{\Gamma(n+a_j-\frac{p_j}{q_j} +b_j \eps)}
     {\Gamma(n+c_j-\frac{p_j}{q_j} +d_j \eps)},
\eq
where $n$, $a_j$ and $b_j$ are integers.
Here the same fraction $p_j/q_j$ occurs in the numerator and the denominator.
In products of these ratios, different fractions are allowed, like
for example in
\bq
\frac{\Gamma(n+a_1-\frac{1}{2} +b_1 \eps)}
     {\Gamma(n+c_1-\frac{1}{2} +d_1 \eps)}
\frac{\Gamma(n+a_2-\frac{2}{3} +b_2 \eps)}
     {\Gamma(n+c_2-\frac{2}{3} +d_2 \eps)}.
\eq
The restriction to ratios of the form (\ref{restrictionnumdenom}) ensures 
that the prefactors $\Gamma(n+1-p/q)$ from the expansion of the Gamma function
in eq. (\ref{Gamma_rational})
cancel between the numerator and the denominator.
The restriction in eq. (\ref{restrictionnumdenom}) will be relaxed in 
sect. \ref{sec:binomial} and sect. \ref{sec:invbinomial}.
We now consider sums of the type A and B, where we allow the substitutions
\bq
a_j \rightarrow a_j -\frac{p_j}{q_j},
 & &
c_j \rightarrow c_j -\frac{p_j}{q_j},
 \nonumber \\
a_j' \rightarrow a_j' -\frac{p_j'}{q_j'},
 & &
c_j' \rightarrow c_j' -\frac{p_j'}{q_j'},
 \nonumber \\
c \rightarrow c - \frac{p_0}{q_0},
 & &
c' \rightarrow c' - \frac{p_0'}{q_0'},
\eq
in eqs. (\ref{algo_A}) - (\ref{algo_B}).
For example, a sum of type A involving rational numbers is of the form
\bq
\label{type_AA}
     \sum\limits_{i=1}^n \frac{x^i}{(i+c-\frac{p_0}{q_0})^m} 
       \frac{\Gamma(i+a_1-\frac{p_1}{q_1}+b_1\varepsilon)}
            {\Gamma(i+c_1-\frac{p_1}{q_1}+d_1\varepsilon)} ...
       \frac{\Gamma(i+a_k-\frac{p_k}{q_k}+b_k\varepsilon)}
            {\Gamma(i+c_k-\frac{p_k}{q_k}+d_k\varepsilon)}
       S(i+o,m_1,...,x_1,...),
\eq
where $c_0$ is a non-negative integer, $o$ and all $a_j$, $c_j$ are integers.
When dealing with rational numbers it is more convenient to work with $S$-sums instead
of $Z$-sums and we replaced the $Z$-sum in eq. (\ref{algo_A}) by a $S$-sum in eq. (\ref{type_AA}).
Due to the conversion algorithm the two formulations are equivalent.

The algorithms for the expansion in $\eps$ of these sums starts
by reducing the offsets $o$ and $o'$ in the subsums $S(i+o;m_1,...;x_1,...)$ and
$S(n-i+o';m_1',...;x_1',...)$ to zero.
Then, using the identity 
\bq
\Gamma(x+1) & = & x \Gamma(x) 
\eq
for the Gamma function, the ratios of the Gamma functions are 
brought to the form
\bq
\frac{\Gamma(i+1-\frac{p_j}{q_j} +b_j \eps)}
     {\Gamma(i+1-\frac{p_j}{q_j} +d_j \eps)}
 \;\;\;\mbox{or}\;\;\;
\frac{\Gamma(n-i+1-\frac{p_j'}{q_j'} +b_j' \eps)}
     {\Gamma(n-i+1-\frac{p_j'}{q_j'} +d_j' \eps)}.
\eq
They are then expanded in $\eps$, using eq. (\ref{Gamma_rational}).
This yields $S$-sums with upper summation limit $(q_j \cdot i)$
or $(q_j' \cdot (n-i))$.
Now, let $q = \mbox{lcm}(q_0,q_1,...,q_k)$ be the least common
multiple of $q_0$, $q_1$, ..., $q_k$ 
and let
$q' = \mbox{lcm}(q_0',q_1',...,q_k')$ be the least common
multiple of $q_0'$, $q_1'$, ..., $q_k'$
Using the refinement algorithm we can convert any occuring $S$-sum $S(q_j i; ...)$ to
$S$-sums with upper summation limit $(q \cdot i)$.
Similar any $S$-sum $S(q_j' (n-i); ...)$ is converted to $S$-sums
$S(q' (n-i);...)$.
Products of $S$-sums are then converted into single $S$-sums
with the help of the multiplication algorithm.
After partial fractioning one arrives at the following forms
\bq
& & \mbox{Type A:} \;\;\; \;\;\;
\sum\limits_{i=1}^n \frac{x^i}{\left(i+c-\frac{p_0}{q_0}\right)^m} S(qi; ...),
 \nonumber \\
& & \mbox{Type B:} \;\;\; \;\;\;
\sum\limits_{i=1}^{n-1} \frac{x^i}{\left(i+c-\frac{p_0}{q_0}\right)^m} S(qi; ...) S\left( q'(n-i); ... \right),
\eq
where $q_0$ divides $q$.
For sums of type A one then reduces the offset $c$ to zero and one arrives 
at sums of the form
\bq
     \sum\limits_{i=1}^n \frac{x^i}{(i-\frac{p_0}{q_0})^m} 
       S(q i, ...),
\eq
Using eq. (\ref{eq_roots_2}) this sum can be written
as
\bq
     \sum\limits_{i=1}^n \frac{x^i}{(i-\frac{p_0}{q_0})^m} 
       S(q i, ...),
 & = &
 q^m x^{p/q} \sum\limits_{i=1}^{q n} \delta_{p,q}(i) 
               \frac{\left( x^{1/q} \right)^i}{i^m} S(i+p;...)
 \nonumber \\
 & = &
q^{m-1} \sum\limits_{l=0}^{q-1}
        \left( r_q^l x^{1/q} \right)^p
        \sum\limits_{i=1}^{q n} 
           \frac{\left( r_q^l x^{1/q} \right)^i}{i^m} S(i+p;...),
\eq
where $p = p_0 q/ q_0$ is an integer with $0 \le p < q$.
Finally, reducing the offset $p$ of the subsum $S(i+p;...)$ to zero,
one arrives at $S$-sums with upper summation limit $(q n)$.

For sums of type B one first refines the subsums
$S(qi; ...)$ and  $S\left( q'(n-i); ... \right)$ to
$S(\hat{q} i; ...)$ and  $S\left( \hat{q}(n-i); ... \right)$, respectively.
Here $\hat{q} = \mbox{lcm}(q,q')$ is the least common multiple of $q$ and $q'$.
The next step consists in rewriting
\bq
\lefteqn{
\sum\limits_{i=1}^{n-1} \frac{x^i}{\left(i+c-\frac{p_0}{q_0}\right)^m} S(qi; ...) S\left( q(n-i); ... \right)
 = } & &
 \nonumber \\
 & &
 q^m x^{p/q} \sum\limits_{i=1}^{q n-q} \delta_{p,q}(i) 
   \frac{ \left(x^{1/q}\right)^i}{\left(i+ q c\right)^m} S(i+p; ...) S\left( q n-i-p; ... \right),
\eq
which brings us back to the integer case.

This completes the necessary modifications for the extension towards the expansion around
rational numbers for the algorithm A and B.


\section{Binomial sums and generalizations}
\label{sec:binomial}

Here we study sums of the form
\bq
\label{genbinomsum}
 \frac{1}{\Gamma\left(1-\frac{p}{q}\right)} \sum\limits_{n=1}^\infty 
 \frac{\Gamma\left(n+1-\frac{p}{q}\right)}{\Gamma\left(n+1\right)} \frac{x_0^n}{n^{m_0}}
   S(n;m_1,...,m_k;x_1,...,x_k).
\eq
This relaxes the condition (\ref{restrictionnumdenom}) and allows
one unbalanced fraction
\bq
 \frac{\Gamma\left(n+1-\frac{p}{q}\right)}{\Gamma\left(n+1\right)}.
\eq
A special case of the form in eq. (\ref{genbinomsum}) are binomial sums
\bq
\label{binomsum}
 \sum\limits_{n=1}^\infty 
\left(
 \begin{array}{c} 
 2 n \\ n \\
 \end{array}
\right) 
   \frac{z^n}{n^{m_0}}
   S(n;m_1,...,m_k;x_1,...,x_k),
\eq
which are obtained by setting $p=1$, $q=2$, $x_0=4z$ in eq. (\ref{genbinomsum}) and by noting the identity
\bq
\left(
 \begin{array}{c} 
 2 n \\ n \\
 \end{array}
\right)
 & = & 
 4^n \frac{\Gamma\left(n+\frac{1}{2}\right)}{\Gamma\left(\frac{1}{2}\right) \Gamma(n+1)}.
\eq
Binomial sums have been studied in 
\cite{Fleischer:1998nb,Jegerlehner:2002em}.
The evaluation of the sums in eq. (\ref{genbinomsum})
follows the same pattern of steps as the evaluation of the conjugation in eq. (\ref{manipconj}), as given
in ref. \cite{Moch:2001zr}.
Due to eq. (\ref{convertS2B}) we may replace in eq. (\ref{genbinomsum}) the $S$-sum $S(n;...)$ with
$S$-sums $S(\infty;...)$  and $B$-sums $B(n;\infty;...)$ and it is sufficient
to study sums of the form
\bq
\label{eqexplicitB}
\lefteqn{
 \frac{1}{\Gamma\left(1-\frac{p}{q}\right)} \sum\limits_{n=1}^\infty 
 \frac{\Gamma\left(n+1-\frac{p}{q}\right)}{\Gamma\left(n+1\right)} \frac{x_0^n}{n^{m_0}}
   B(n;\infty;m_1,...,m_k;x_1,...,x_k)
 = } \nonumber \\
 & &
   - \left( {\bf x_k}^+ \right)^{m_k} \left( {\bf x_{k-1}}^+ \right)^{m_{k-1}} ... 
     \left( {\bf x_1}^+ \right)^{m_1} \left( {\bf x_0}^+ \right)^{m_0}
      \frac{x_k}{1-x_k} \frac{x_{k-1}x_k}{1-x_{k-1}x_k} ...
     \frac{x_1 ... x_k}{1-x_1 ... x_k}
 \nonumber \\
 & &
 \times
 \left[ 1 - \left(1-x_0 x_1 ... x_k\right)^{-\left(1-\frac{p}{q}\right)} \right].
\eq
To derive the r.h.s of eq. (\ref{eqexplicitB}) we first introduced the raising operators ${\bf x_+}$, 
then performed explicitly all geometric sums from $B(n;\infty;0,...,0;x_1,...,x_k)$ and finally performed
the remaining sum with the help of the hypergeometric summation formula
\bq
 \frac{1}{\Gamma\left(1-\frac{p}{q}\right)} \sum\limits_{n=1}^\infty
 \frac{\Gamma\left(n+1-\frac{p}{q}\right)}{\Gamma\left(n+1\right)} x^n
 & = & 
 - \left[ 1 - \left(1-x\right)^{-\left(1-\frac{p}{q}\right)} \right].
\eq
We then perform succesivly the integrations corresponding to the raising operators.
Due to the appearance of rational numbers in the exponent, the set in eq. (\ref{raisint})
is no longer sufficient and has to be supplemented with additional equations.
Nevertheless, the principle stays the same: Each integration preserves the structure, such that multiple
integrations can be performed iteratively.
To reduce $m_0$ we have
\bq
\lefteqn{
{\bf x_0^+} \left[ 1 - \left(1-x_0x_1\right)^{-\left(1-\frac{p}{q}\right)} \right]
 = 
 q \sum\limits_{n=1}^\infty
   \frac{1}{n} \left[ \delta_{0,q}(n) - \delta_{q-p,q}(n) \right]
                 \left[ 1- \left( 1 - x_0 x_1 \right)^{\frac{n}{q}} \right],
} & &
 \nonumber \\
\lefteqn{
{\bf x_0^+} \delta_{q-p,q}(n) \left[ 1- \left( 1 - x_0 x_1 \right)^{\frac{n}{q}} \right] 
 =
 q \delta_{q-p,q}(n)
 \sum\limits_{i=1}^\infty
   \frac{1}{i} \left[ \delta_{0,q}(i) - \delta_{q-p,q}(i) \right] 
                        \left[ 1- \left( 1 - x_0 x_1 \right)^{\frac{i}{q}} \right]
} \nonumber \\
 & &
 + q \delta_{q-p,q}(n)
   \sum\limits_{i=1}^n \frac{1}{i}
    \delta_{q-p,q}(i) \left[ 1- \left( 1 - x_0 x_1 \right)^{\frac{i}{q}} \right].
 \hspace*{70mm}
\eq
To reduce $m_1$, $m_2$, ..., $m_k$ we have in the general case $x_0 \neq 1$:
\bq
\label{redm1m2mk}
\lefteqn{
{\bf x_1^+} \frac{x_1 x_2}{1-x_1x_2} \left[ 1 - \left(1-x_0x_1x_2\right)^{-\left(1-\frac{p}{q}\right)} \right]
 = 
 - q \left( 1 -x_0 \right)^{-\left(1-\frac{p}{q}\right)}
} & &
 \\
 & &
 \times 
  \left\{ \sum\limits_{n=1}^\infty
           \frac{1}{n} 
           \left[ \delta_{0,q}(n) - \delta_{q-p,q}(n) \right]
           \left( \frac{1}{1-x_0} \right)^{\frac{n}{q}}
           \left[ 1 - \left( 1 -x_0x_1x_2\right)^{\frac{n}{q}} \right] 
  \right\}
 \nonumber \\
 & & 
 + \left[ 1 - \left( 1-x_0\right)^{-\left(1-\frac{p}{q}\right)} \right]
   \sum\limits_{n=1}^N \frac{1}{n} \left( x_1 x_2 \right)^n
 + {\bf x_1^+} \frac{x_1 x_2}{1-x_1x_2} \left( x_1 x_2 \right)^N 
       \left[ 1 - \left(1-x_0\right)^{-\left(1-\frac{p}{q}\right)} \right],
 \nonumber \\
\lefteqn{
{\bf x_1^+} \delta_{q-p,q}(n) \frac{x_1 x_2}{1-x_1x_2} \left[ 1 - \left(1-x_0x_1x_2\right)^{\frac{n}{q}} \right]
 = 
 - q \left( 1 -x_0 \right)^{\frac{n}{q}}
   \delta_{q-p,q}(n)
} & &
 \nonumber \\
 & &
 \times
  \left\{ \sum\limits_{i=1}^\infty
           \frac{1}{i} 
           \left[ \delta_{0,q}(i) - \delta_{q-p,q}(i) \right]
           \left( \frac{1}{1-x_0} \right)^{\frac{i}{q}}
           \left[ 1 - \left( 1 -x_0x_1x_2\right)^{\frac{i}{q}} \right] 
 \right.
 \nonumber \\
 & &
 \left.
        + \sum\limits_{i=1}^n
           \frac{1}{i} \delta_{q-p,q}(i)
           \left( \frac{1}{1-x_0} \right)^{\frac{i}{q}}
           \left[ 1 - \left( 1 -x_0x_1x_2\right)^{\frac{i}{q}} \right] 
  \right\}
 \nonumber \\
 & & 
 + \delta_{q-p,q}(n)
   \left[ 1 - \left( 1-x_0\right)^{\frac{n}{q}} \right]
   \sum\limits_{i=1}^N \frac{1}{i} \left( x_1 x_2 \right)^i
 + \delta_{q-p,q}(n)
   {\bf x_1^+} \frac{x_1 x_2}{1-x_1x_2} \left( x_1 x_2 \right)^N 
       \left[ 1 - \left(1-x_0\right)^{\frac{n}{q}} \right].
 \nonumber 
\eq
In the special case $x_0=1$ we have instead:
\bq
\label{redm1m2mkx01}
\lefteqn{
{\bf x_1^+} \frac{x_1 x_2}{1-x_1x_2} \left[ 1 - \left(1-x_1x_2\right)^{-\left(1-\frac{p}{q}\right)} \right]
 = 
 \frac{1}{1-\frac{p}{q}} \left[ 1 - \left(1 -x_1x_2 \right)^{-\left(1-\frac{p}{q}\right)} \right]
} & & \\ 
 & &
 \hspace*{60mm}
 + 
   \sum\limits_{n=1}^N \frac{1}{n} \left( x_1 x_2 \right)^n
 + {\bf x_1^+} \frac{x_1 x_2}{1-x_1x_2} \left( x_1 x_2 \right)^N,
 \hspace*{25mm}
 \nonumber \\
\lefteqn{
{\bf x_1^+} \frac{x_1 x_2}{1-x_1x_2} \left[ 1 - \left(1-x_1x_2\right)^{\frac{n}{q}} \right]
 =   
 - \frac{q}{n} \left[ 1 - \left(1 -x_1x_2 \right)^{\frac{n}{q}} \right]
 + 
   \sum\limits_{i=1}^N \frac{1}{i} \left( x_1 x_2 \right)^i
 + {\bf x_1^+} \frac{x_1 x_2}{1-x_1x_2} \left( x_1 x_2 \right)^N.
} & &
 \nonumber 
\eq
Eqs. (\ref{redm1m2mk}) and (\ref{redm1m2mkx01}) introduce an arbitrary integer $N$. 
It is worth to note that the eqs. (\ref{redm1m2mk}) and (\ref{redm1m2mkx01})
hold for any integer $N$.
We will take the limit $N \rightarrow \infty$ in the end.
It is clear that in this limit terms of the form
\bq
\left( {\bf x^+} \right)^m \frac{x}{1-x} x^N & = & \sum\limits_{i=N+1}^\infty \frac{x^i}{i^m}
\eq
can be neglected.
Some examples are:
\bq
\lefteqn{
\frac{1}{\Gamma\left(\frac{1}{2}\right)}
 \sum\limits_{n=1}^\infty 
 \frac{\Gamma\left(n+\frac{1}{2}\right)}{\Gamma(n+1)}
 \frac{x^n}{n}
 =  
 2 \ln 2 - 2 \ln\left(1+\sqrt{1-x}\right),
} & &
 \nonumber \\
\lefteqn{
\frac{1}{\Gamma\left(\frac{1}{2}\right)}
 \sum\limits_{n=1}^\infty 
 \frac{\Gamma\left(n+\frac{1}{2}\right)}{\Gamma(n+1)}
 \frac{x^n}{n^2}
 =  
 2 \left[ 
          \ln 2 \ln \left( 1+ \sqrt{1-x} \right)
        + \ln 2 \ln \left( 1- \sqrt{1-x} \right)
        - \mbox{Li}_{11}\left(-\sqrt{1-x},1\right)
 \right.
 } & & \nonumber \\
 & & \left.
        - \mbox{Li}_{11}\left(\sqrt{1-x},-1\right)
        - \mbox{Li}_{2}\left(-1\right)
        - \left( \ln 2 \right)^2
   \right],
 \hspace*{80mm}
 \nonumber \\
\lefteqn{
\frac{1}{\Gamma\left(\frac{1}{2}\right)}
 \sum\limits_{n=1}^\infty 
 \sum\limits_{j=1}^n 
 \frac{\Gamma\left(n+\frac{1}{2}\right)}{\Gamma(n+1)}
 \frac{x^n}{n}
 \frac{y^j}{j}
 = 
  2 \; \mbox{Li}_{11}\left( \sqrt{1-x}, -\sqrt{\frac{1-xy}{1-x}}\right)
  + 2 \; \mbox{Li}_{11}\left( -\sqrt{1-x}, \sqrt{\frac{1-xy}{1-x}}\right)
} & & 
 \nonumber \\
 & &
  - 2 \; \mbox{Li}_{11}\left( \sqrt{1-x}, -\frac{1}{\sqrt{1-x}}\right)
  - 2 \; \mbox{Li}_{11}\left( -\sqrt{1-x}, \frac{1}{\sqrt{1-x}}\right)
  + 4 \; \mbox{Li}_{2}\left(-\sqrt{1-xy}\right)
 \nonumber \\
 & &
  - 4 \; \mbox{Li}_{2}(-1)
  + 2 \ln \frac{1+\sqrt{1-x}}{1-\sqrt{1-x}} 
      \ln \frac{1+\sqrt{\frac{1-x y}{1-x}}}{1+\frac{1}{\sqrt{1-x}}}.
\eq


\section{Inverse binomial sums and generalizations}
\label{sec:invbinomial}

In this section I consider sums with one unbalanced rational number in the denominator.
Examples of these type are sums of the form
\bq
\label{geninvbinomsum}
\lefteqn{
 \Gamma\left(1-\frac{p}{q}\right) \sum\limits_{n_1=1}^\infty 
 \frac{\Gamma\left(n_1+a\right)}{\Gamma\left(n_1+b-\frac{p}{q}\right)}
   x_1^{n_1}
} & & \nonumber \\
 & & 
 \times
 \sum\limits_{n_2=1}^{n_1-1} \frac{x_2^{n_2}}{\left(n_1-n_2\right)^{m_1}}
 ...
 \sum\limits_{n_{k-1}=1}^{n_{k-2}-1} \frac{x_{k-1}^{n_{k-1}}}{\left(n_{k-2}-n_{k-1}\right)^{m_{k-2}} }
 \sum\limits_{n_k=1}^{n_{k-1}-1} \frac{x_k^{n_k}}{\left(n_{k-1}-n_k\right)^{m_{k-1}} n_k^{m_k}}.
\eq
Note the slightly different structure of $(n_{l+1}-n_l)$ in the denominators of the subsums.
Only in the innermost sum a factor $n_k$ appears in the denominator.
Changing the summation variables according to
\bq
n_1 = j_1 + j_2 + ... + j_k, 
 \;\;\;
n_2 =       j_2 + ... + j_k, 
 \;\;\;
...,
 \;\;\;
n_k & = & j_k,
\eq
this sum is equivalent to
\bq
\label{geninvbinomsum2}
 \Gamma\left(1-\frac{p}{q}\right) 
 \sum\limits_{j_1=1}^\infty 
 ...
 \sum\limits_{j_k=1}^\infty 
 \frac{\Gamma\left(j_1+...+j_k+a\right)}{\Gamma\left(j_1+...+j_k+b-\frac{p}{q}\right)}
 \frac{x_1^{j_1}}{j_1^{m_1}}
 \frac{\left( x_1 x_2 \right)^{j_2}}{j_2^{m_2}}
 ...
 \frac{\left( x_1 ... x_k\right)^{j_k}}{j_k^{m_k}}.
\eq
A special case of this form are inverse binomial sums
\bq
\label{invbinomsum}
 \sum\limits_{n=1}^\infty 
\frac{1}{\left(
                \begin{array}{c} 
                2 n \\ n \\
                \end{array}
         \right)} 
   \frac{z^n}{n^{m_1}},
\eq
which are obtained by setting $p=1$, $q=2$, $a=b=1$, $k=1$ and $x_1=z/4$ in eq. (\ref{geninvbinomsum2}).
Inverse binomial sums have been studied in 
\cite{Ogreid:1998bx,Fleischer:1998nb,Borwein:2000et,Davydychev:2000na,Kalmykov:2000qe,Davydychev:2003mv}
and have been related in the literature to log-sine integrals
\cite{Fleischer:1999mp,Davydychev:1999mq}.
Here I follow a different path for the evaluation of inverse binomial sums.
The method presented here is closely connected to the transformation
\bq
x' & = & \frac{-x}{1-x},
\eq
whose inverse transformation is again given by
\bq
x & = & \frac{-x'}{1-x'}.
\eq
To start with the evaluation of sums of type (\ref{geninvbinomsum2}) one replaces all factors
$1/j$ by
\bq
\frac{1}{j} & = & \lim\limits_{\lambda \rightarrow 0} \frac{\Gamma(j+\lambda)}{\Gamma(j+1+\lambda)}.
\eq
The original sum is recovered as the $\lambda \rightarrow 0$ limit of the regularized sum.
The introduction of the regularization in $\lambda$ allows us to extend the lower summation boundary
of all sums from $1$ to $0$.
This can be done recursively according to
\bq
\sum\limits_{j=1}^\infty f(j) = - f(0) + \sum\limits_{j=0}^\infty f(j).
\eq
$f(0)$ corresponds to a sum of lower depth.
We therefore deal with sums of the type
\bq
\lefteqn{
 \Gamma\left(1-\frac{p}{q}\right) 
 \sum\limits_{j_1=1}^\infty 
 ...
 \sum\limits_{j_k=1}^\infty 
 \frac{\Gamma\left(j_1+...+j_k+a\right)}{\Gamma\left(j_1+...+j_k+b-\frac{p}{q}\right)}
} & &
 \nonumber \\
 & & 
  x_1^{j_1} \left( x_1 x_2 \right)^{j_2} ... \left( x_1 ... x_k\right)^{j_k}
  \left[ \frac{\Gamma\left(j_1+\lambda\right)}{\Gamma\left(j_1+1+\lambda\right)} \right]^{m_1}
  ...
  \left[ \frac{\Gamma\left(j_k+\lambda\right)}{\Gamma\left(j_k+1+\lambda\right)} \right]^{m_k}.
\eq
For these sums we have the following integral representation:
\bq
\label{invbinintrep1}
\lefteqn{
 \frac{\Gamma\left(1-\frac{p}{q}\right)}{\Gamma\left(b-a-\frac{p}{q}\right)}
 \int\limits_0^1 du u^{a-1} \left( 1 - u \right)^{b-a-\frac{p}{q}-1}
 \int\limits_0^1 dt_1 t_1^{\lambda-1} 
  ...
 \int\limits_0^1 dt_{i_k} t_{i_k}^{\lambda-1} 
} & &
 \nonumber \\
 & &
 \left( 1 - u t_1 ... t_{i_1} x_1 \right)^{-1}
 \left( 1 - u t_{i_1+1} ... t_{i_2} x_1 x_2 \right)^{-1}
 ...
 \left( 1 - u t_{i_{k-1}+1} ... t_{i_k} x_1 ... x_k \right)^{-1},
\eq
where
\bq 
i_1=m_1, \;\;\; i_2=m_1+m_2, \;\;\; ..., \;\;\; i_k=m_1+...+m_k.
\eq
For the integration variable $u$ we now perform the substitution
\bq
\label{subsvarint}
u \rightarrow 1-u,
\eq
and we obtain for eq. (\ref{invbinintrep1})
\bq
\lefteqn{
 \frac{\Gamma\left(1-\frac{p}{q}\right) \Gamma\left(a \right)}{\Gamma\left(b-a-\frac{p}{q}\right)}
 \sum\limits_{j_1=0}^\infty ... \sum\limits_{j_k=0}^\infty
 \frac{\Gamma\left(j_1+...+j_k+b-a-\frac{p}{q}\right)}{\Gamma\left(j_1+...+j_k+b-\frac{p}{q}\right)}
 \int\limits_0^1 dt_1 t_1^{\lambda-1} 
  ...
 \int\limits_0^1 dt_{i_k} t_{i_k}^{\lambda-1} 
} & &
 \nonumber \\
 & &
  \left( - t_1 ... t_{i_1} x_1 \right)^{j_1} \left( 1 - t_1 ... t_{i_1} x_1 \right)^{-j_1-1}
  ...
  \left( - t_{i_{k-1}+1} ... t_{i_k} x_1 ... x_k \right)^{j_k} 
   \left( 1 - t_{i_{k-1}+1} ... t_{i_k} x_1 ... x_k \right)^{-j_k-1}
 \nonumber \\
& = &
 \frac{\Gamma\left(1-\frac{p}{q}\right) \Gamma\left(a \right)}{\Gamma\left(b-a-\frac{p}{q}\right)}
 \sum\limits_{n_1=0}^\infty ... \sum\limits_{n_k=0}^{n_{k-1}}
 \frac{\Gamma\left(n_1+b-a-\frac{p}{q}\right)}{\Gamma\left(n_1+b-\frac{p}{q}\right)}
 \int\limits_0^1 dt_1 t_1^{\lambda-1} 
  ...
 \int\limits_0^1 dt_{i_k} t_{i_k}^{\lambda-1} 
 \nonumber \\
 & &
  \left( - t_1 ... t_{i_1} x_1 \right)^{n_1-n_2} \left( 1 - t_1 ... t_{i_1} x_1 \right)^{-n_1+n_2-1}
 \nonumber \\
 & &
  ...
  \left( - t_{i_{k-2}+1} ... t_{i_{k-1}} x_1 ... x_{k-1} \right)^{n_{k-1}-n_k} 
   \left( 1 - t_{i_{k-2}+1} ... t_{i_{k-1}} x_1 ... x_{k-1} \right)^{-n_{k-1}+n_k-1}
 \nonumber \\
 & &
  \left( - t_{i_{k-1}+1} ... t_{i_k} x_1 ... x_k \right)^{n_k} 
   \left( 1 - t_{i_{k-1}+1} ... t_{i_k} x_1 ... x_k \right)^{-n_k-1}.
\eq
Note that the substitution eq. (\ref{subsvarint}) induces the 
transformation $x \rightarrow -x/(1-x)$ as follows:
\bq
\left[ 1 - \left(1-u\right) x \right]^{-c} 
 & = & 
 \left( 1 - x \right)^{-c} \left[ 1 - \left( \frac{-x}{1-x} \right) u \right]^{-c}
\eq
As a sideremark it is worth to note that as special case one obtains by this procedure
the transformation of the hypergeoemetric function
\bq
{}_2F_1(a,b;c;x) & = & \left(1-x\right)^{-a} {}_2F_1\left(a,c-b;c;\frac{-x}{1-x}\right)
 = \left(1-x\right)^{-b} {}_2F_1\left(c-a,b;c;\frac{-x}{1-x}\right). \nonumber
\eq
The purpose of the substitution in eq. (\ref{subsvarint}) 
is to change the ratio of Gamma functions from
\bq
\frac{\Gamma\left(j_1+...+j_k+a\right)}{\Gamma\left(j_1+...+j_k+b-\frac{p}{q}\right)}
\;\;\;
\mbox{to}
\;\;\;
\frac{\Gamma\left(j_1+...+j_k+b-a-\frac{p}{q}\right)}{\Gamma\left(j_1+...+j_k+b-\frac{p}{q}\right)}.
\eq
In the last form, the rational number $p/q$ appears both in the numerator and the denominator
and can therefore be treated with the methods discussed in sect. \ref{sec:algos}.
It remains to perform the integration over $t_1$, ..., $t_{i_k}$.
These integration are recursively done according to the formula
\bq
\int\limits_0^1 dt_1 t_1^{a-1} \left( - t_0 t_1 x \right)^n
  \left( 1- t_0 t_1 x \right)^{-n-c}
 & = & 
 \frac{\Gamma\left(n+a\right)}{\Gamma\left(n+c\right)}
 \sum\limits_{k=n}^\infty  \frac{\Gamma\left(k+c\right)}{\Gamma\left(k+1+a\right)}
 \left( - t_0 x \right)^k
  \left( 1- t_0 x \right)^{-k-c}.
 \nonumber 
\eq
As a net result we obtain a rooted tree with side-branches, which can be expanded in $\lambda$ and converted
to nested sums with the help of the algorithms discussed in
ref. \cite{Moch:2001zr} and in sect. \ref{sec:algos}.
In the final result all poles in $\lambda$ cancel and one can extract the $\lambda^0$-term, which yields
the evaluation of the inverse binomial sum eq. (\ref{geninvbinomsum}).
Some examples are:
\bq
\label{examplesinvbin}
\lefteqn{
\Gamma\left(\frac{1}{2}\right)
 \sum\limits_{n=1}^\infty 
 \frac{\Gamma(n+1)}{\Gamma\left(n+\frac{1}{2}\right)}
 \frac{x^n}{n}
 =  
 \chi \ln \frac{1-\chi}{1+\chi}, }
 & &
 \nonumber \\
\lefteqn{
\Gamma\left(\frac{1}{2}\right)
 \sum\limits_{n=1}^\infty 
 \frac{\Gamma(n+1)}{\Gamma\left(n+\frac{1}{2}\right)}
 \frac{x^n}{n^2}
 =  
 - \mbox{Li}_{11}\left(\chi,1\right)
 + \mbox{Li}_{11}\left(\chi,-1\right)
 - \mbox{Li}_{11}\left(-\chi,1\right)
 + \mbox{Li}_{11}\left(-\chi,-1\right),
} & &
 \nonumber \\
\lefteqn{
\Gamma\left(\frac{1}{2}\right)
 \sum\limits_{n=1}^\infty \sum\limits_{j=1}^n 
 \frac{\Gamma(n+1)}{\Gamma\left(n+\frac{1}{2}\right)}
 \frac{x^n}{n} \frac{y^j}{j} 
 =  
        - \mbox{Li}_{11}\left(\upsilon,1\right)
        - \mbox{Li}_{11}\left(-\upsilon,1\right)
        + \mbox{Li}_{11}\left(\upsilon,-1\right)
        + \mbox{Li}_{11}\left(-\upsilon,-1\right)
 } & &
 \nonumber \\
 & &
 - \chi
   \left[
          \mbox{Li}_{11}\left(\chi, \frac{\upsilon}{\chi} \right)
        + \mbox{Li}_{11}\left(\chi, -\frac{\upsilon}{\chi} \right)
        - \mbox{Li}_{11}\left(-\chi, \frac{\upsilon}{\chi} \right)
        - \mbox{Li}_{11}\left(-\chi, -\frac{\upsilon}{\chi} \right)
   \right] 
 \nonumber \\
 & &
  - \chi \ln\left(1-xy\right) 
   \ln \frac{1-\chi}{1+\chi},
 \hspace*{100mm}
\eq
where $\chi = \sqrt{-x/(1-x)}$ and $\upsilon = \sqrt{-xy/(1-xy)}$.


\section{Applications}
\label{sec:appl}

In this section I give some applications of the techniques described in the previous sections
relevant to the calculation of Feynman loop integrals with massive particles
or to phase space integrals.
The first example concerns a one-loop triangle with a uniform internal mass
$m_1=m_2=m_3=m$ and two vanishing external momenta $p_1^2=p_2^2=0$.
The third external momentum $p_3^2=p^2$ is kept arbitrary.
Such diagrams are for example relevant to Higgs physics.
Specific examples are Higgs production via $g g \rightarrow H$ or Higgs decay into two
photons $H \rightarrow \gamma \gamma$. In both cases the interaction
proceeds via an internal top quark loop.
It is well known that the corresponding integral 
in dimensional regularization 
is proportional to
\cite{Boos:1991rg,Davydychev:2000na}
\bq
{}_3F_2\left( 1, 1, 1+\eps; \; \frac{3}{2}, 2; \; \frac{p^2}{4 m_{top}^2} \right).
\eq
Here $\eps=(4-D)/2$ denotes the deviation from four space-time dimensions.
The expansion of the hypergeometric function to sufficient high order in $\eps$ is 
a non-trivial task \cite{Davydychev:2000na,Davydychev:2003mv}.
The expansion can be accomplished systematically with the methods presented here.
First one notices that there is one unbalanced half-integer number in the denominator.
With the results of sect. \ref{sec:invbinomial} one can show that
\bq
\lefteqn{
{}_3F_2\left( 1+a_1\eps, 1+a_2\eps, 1+a_3\eps; \; \frac{3}{2}+b\eps, C+c\eps; \; z \right)
 =  
 \left( 1 - z \right)^{-1-a_1\eps}
} & & 
 \nonumber \\
 & &
 \frac{\Gamma\left(\frac{3}{2}+b\eps\right) \Gamma\left(C+c\eps\right)}
      {\Gamma\left(1+a_1\eps\right) \Gamma\left( 1+a_3\eps\right)
       \Gamma\left(\frac{1}{2}+\left(b-a_2\right)\eps\right) 
       \Gamma\left( C-1+\left(c-a_3\right)\eps\right)}
 \sum\limits_{n=0}^\infty 
   \frac{\Gamma\left(n+1+a_1\eps\right)}{\Gamma\left(n+C+c\eps\right)}
   \left( \frac{-z}{1-z} \right)^n
 \nonumber \\
 & &
   \sum\limits_{j=0}^n
     \frac{\Gamma\left(j+1+a_3\eps\right)}{\Gamma\left(j+1\right)}
     \frac{\Gamma\left(j+\frac{1}{2}+\left(b-a_2\right)\eps\right)}
          {\Gamma\left(j+\frac{3}{2}+b\eps\right)}
     \frac{\Gamma\left(n-j+C-1+\left(c-a_3\right)\eps\right)}
          {\Gamma\left(n-j+1\right)}.
\eq
This yields a nested sum where all half-integer numbers are balanced in the numerator
and in the denominator.
This expression can now be expanded systematically in $\eps$ with the algorithms described
in sect. \ref{sec:algos}.
For the example mentioned above one finds
\bq
\lefteqn{
{}_3F_2\left( 1, 1, 1+\eps; \; \frac{3}{2}, 2; \; z \right)
 =  
- \frac{1}{2z} \left(1-z\right)^{-\eps}
} & &
 \nonumber \\
 & &
 \left\{
  \mbox{Li}_{11}\left( \sqrt{\frac{-z}{1-z}}, 1 \right)
  + \mbox{Li}_{11}\left( -\sqrt{\frac{-z}{1-z}}, 1 \right)
  - \mbox{Li}_{11}\left( \sqrt{\frac{-z}{1-z}}, -1 \right)
  - \mbox{Li}_{11}\left( -\sqrt{\frac{-z}{1-z}}, -1 \right)
 \right. \nonumber \\
 & & \left.
  + 2 \eps
    \left[
  \mbox{Li}_{111}\left( \sqrt{\frac{-z}{1-z}}, 1, 1 \right)
  + \mbox{Li}_{111}\left( -\sqrt{\frac{-z}{1-z}}, 1, 1 \right)
  - \mbox{Li}_{111}\left( \sqrt{\frac{-z}{1-z}}, -1, -1 \right)
 \right. \right. \nonumber \\
 & & \left. \left.
  - \mbox{Li}_{111}\left( -\sqrt{\frac{-z}{1-z}}, -1, -1 \right)
    \right]
  + {\cal O}\left(\eps^2\right)
 \right\}.
\eq
As a second example I discuss briefly phase space integrals.
Analytic results for phase space integrals are needed for example in the calculation
of higher order corrections for jet observables as integrals over approximation terms
within the subtraction method \cite{Catani:1997vz,Phaf:2001gc,Weinzierl:2003fx,Weinzierl:2003ra}.
An example is the integral over a subtraction term, which approximates the emission
of a soft gluon from a heavy quark pair.
This integral is given by
\cite{Phaf:2001gc}
\bq
{\cal V}_{QQ}(r_0,\eps) & = & 
 \left( \frac{r_0}{2} \right)^{-2\eps}
 \int\limits_0^1 dr r^{-2\eps-1} (1-r)^{-\eps} (1-r_0r)^{-1}
 \int\limits_{-1}^1 ds (1-s^2)^{-\eps} \nonumber \\
 & & 
 \left[
 \left( 2(1-r_0r) -(1-r_0) \right) (1-s_0 s)^{-1} -(1-r_0) (1-s_0 s)^{-2} \right],
\eq
where
\bq
s_0 & = & \sqrt{\frac{r_0 (1-r)}{1 - r_0 r}}.
\eq
This integral corresponds to the following triple sum
\bq
{\cal V}_{QQ}(r_0,\eps) & = & 
 \frac{\Gamma(1-\eps)}{\Gamma(\eps)}
 \left( \frac{r_0}{2} \right)^{-2\eps}
 \sum\limits_{i=0}^\infty
 \sum\limits_{j=0}^\infty
 \sum\limits_{k=0}^\infty
  (-1)^i \left( 1 + (-1)^j \right) r_0^{k+j/2} \nonumber \\
 & &
  \frac{\Gamma(i+j+1) }{\Gamma(i+j+2-\eps)}
  \frac{\Gamma(i+\eps)}{ \Gamma(i+1)}
  \frac{ \Gamma(k-2\eps) \Gamma(j/2+1-\eps) \Gamma(k+j/2) }
       { \Gamma(k+1) \Gamma(j/2) \Gamma(k+j/2+1-3\eps) } \nonumber \\
 & &
  \left[ 2 - (1-r_0) (2+j) \frac{j+2k}{j} \right].
\eq
Using the methods of sect. \ref{sec:algos} this sum can be expanded 
systematically in $\eps$. The result is
\bq
\lefteqn{
{\cal V}_{QQ}(r_0,\eps) =  
  \frac{1}{\eps} \left( 1 - \frac{1}{2} \frac{1+r_0}{\sqrt{r_0}} 
                            \ln \frac{1+\sqrt{r_0}}{1-\sqrt{r_0}} \right) 
  - 2 \ln r_0 
 - \ln^2\left(\frac{1+\sqrt{r_0}}{1-\sqrt{r_0}}\right) 
 +\frac{1}{\sqrt{r_0}}\ln\left(\frac{1+\sqrt{r_0}}{1-\sqrt{r_0}}\right)
} & &
 \nonumber \\
 & & 
 - \frac{1+r_0}{2\sqrt{r_0}} 
   \left[ \mbox{Li}_2\left(\sqrt{r_0}\right) - \mbox{Li}_2\left(-\sqrt{r_0}\right)
          +2\;\mbox{Li}_2\left(\frac{1+\sqrt{r_0}}{2}\right)
          -2\;\mbox{Li}_2\left(\frac{1-\sqrt{r_0}}{2}\right) 
          +\mbox{Li}_2\left(\frac{\sqrt{r_0}-1}{2\sqrt{r_0}}\right)
 \right. \nonumber \\
 & & \left.
          -\mbox{Li}_2\left(\frac{\sqrt{r_0}-1}{\sqrt{r_0}}\right)
          +\mbox{Li}_2\left(\frac{1}{1+\sqrt{r_0}}\right)
          -\mbox{Li}_2\left(\frac{1-\sqrt{r_0}}{1+\sqrt{r_0}}\right)
          -2\ln r_0 \ln\left(\frac{1+\sqrt{r_0}}{1-\sqrt{r_0}}\right)
          +\frac{1}{2} \ln^2 2
 \right. \nonumber \\
 & & \left.
          +\ln2 \ln \frac{\sqrt{r_0}}{1+\sqrt{r_0}}
          + \ln(1-\sqrt{r_0}) \ln \left( \frac{1+\sqrt{r_0}}{\sqrt{r_0}} \right)
          + \frac{1}{2} \ln^2(1+\sqrt{r_0}) 
          - \frac{1}{2} \ln^2(1-\sqrt{r_0})
   \right] 
 \nonumber \\
& & + O(\eps).
\eq


\section{Summary and conclusions}
\label{sec:concl}

In this paper I reported on algorithm which allow the expansion of certain
transcendental functions in a small parameter around rational values.
These algorithms extend the ones for the expansion around integer values
and are based on the manipulation of specific forms of nested sums.
Of particular importance is the case of the expansion around half-integer
values.
This case occurs frequently in the calculation of radiative corrections in quantum
field theories with massive particles.
The methods presented in this paper allow a systematic approach for the calculation
of these integrals.


\begin{appendix}

\section{Log-sine integrals}
\label{append:logsine}

Inverse binomial sums are often expressed in terms of log-sine integrals.
In this appendix I briefly summarize the results from the literature.
The following inverse binomial sum can be evaluated with elementary functions as follows:
\bq
\Gamma\left(\frac{1}{2}\right)
\sum\limits_{n=1}^\infty 
 \frac{\Gamma(n+1)}{\Gamma\left(n+\frac{1}{2}\right)}
 \frac{x^n}{n}
 & = & 
 \frac{2 \sqrt{x} \arcsin\left(\sqrt{x} \right)}{\sqrt{1-x}}
\eq
This result agrees with the one given in eq. (\ref{examplesinvbin}).
In the literature, evaluations of inverse binomial sums of higher weights
are given in terms of log-sine functions
\cite{Ogreid:1998bx,Fleischer:1999mp,Davydychev:1999mq,Fleischer:1998nb,Borwein:2000et,Davydychev:2000na,Kalmykov:2000qe,Davydychev:2003mv}
:
\bq
\Gamma\left(\frac{1}{2}\right)
\sum\limits_{n=1}^\infty  \frac{\Gamma(n+1)}{\Gamma\left(n+\frac{1}{2}\right)}
 \frac{z^n}{n^m}
 & = & 
 - \sum\limits_{j=0}^{m-2} \frac{(-2)^j}{j! (m-2-j)!}
               \left( \ln 4 z \right)^{m-2-j} \mbox{Ls}_{j+2}^{(1)}\left( \theta \right),
\eq
where $\theta = 2 \arcsin \sqrt{z}$ and the log-sine functions are defined by
\bq
\mbox{Ls}_j^{(k)}(\theta) & = & - \int\limits_0^\theta d\theta' \;
                          \left( \theta' \right)^k
                          \ln^{j-k-1} \left| 2 \sin \frac{\theta'}{2} \right|.
\eq
By analytic continuation the log-sine functions are then
related to polylogarithms \cite{Davydychev:2003mv}.
A simple example is given by
\bq
\mbox{Ls}_2^{(0)}(\theta) & = & \mbox{Cl}_2(\theta),
\eq
involving the Clausen function $\mbox{Cl}_2$. 
The Clausen functions $\mbox{Cl}_n$ are given in terms
of polylogarithms by
\bq
\mbox{Cl}_n(\theta) & = & 
 \left\{ \begin{array}{cc}
   \frac{1}{2i} \left[ \mbox{Li}_n\left( e^{i \theta} \right) 
                      -\mbox{Li}_n\left( e^{-i \theta} \right)
                \right], 
   & n \; \mbox{even}, \\
   & \\
    \frac{1}{2} \left[ \mbox{Li}_n\left( e^{i \theta} \right) 
                      +\mbox{Li}_n\left( e^{-i \theta} \right)
                \right], 
   & n \; \mbox{odd}. \\
 \end{array} \right.
\eq

\section{Relations for polylogarithms}
\label{append:polylog}

Multiple polylogarithms of low weight can be expressed in terms of ordinary
logarithms and polylogarithms.
Relations relevant to the examples in this paper are
\bq
\mbox{Li}_1\left(x\right) & = & - \ln\left(1-x\right),
 \nonumber \\
\mbox{Li}_{11}\left(x,1\right) & = &
          \frac{1}{2} \ln^2 \left(1-x \right),
 \nonumber \\
\mbox{Li}_{11}\left(x,-1\right) & = &
 \frac{1}{2} \zeta_2 - \frac{1}{2} \ln^2 2
 - \ln(1-x) \ln(1+x)  + \ln 2 \ln(1+x)
 - \mbox{Li}_2\left(\frac{1+x}{2} \right),
 \nonumber \\
\mbox{Li}_{11}\left(x,y\right) & = &
         \ln\left(1-x\right) \ln\left(1-y\right) 
         +\mbox{Li}_2\left( \frac{-y}{1-y} \right)
         -\mbox{Li}_2\left( \frac{-y(1-x)}{1-y} \right),
 \nonumber \\
\mbox{Li}_{111}\left(x,1,1\right) & = &
 - \frac{1}{6} \ln^3\left(1-x\right),
 \nonumber \\
\mbox{Li}_{111}\left(x,-1,-1\right) & = &
   \frac{1}{2} \zeta_2 \ln 2 
 - \frac{7}{8} \zeta_3 
 - \frac{1}{6} \ln^3 2 
 - \frac{1}{2} \ln\left(1-x\right) \ln^2 \left(1+x\right)
 + \frac{1}{2} \ln 2 \ln^2 \left(1+x\right)
 \nonumber \\
 & &
 - \ln\left(1+x\right) \mbox{Li}_2 \left( \frac{1+x}{2} \right)
 + \mbox{Li}_3 \left( \frac{1+x}{2} \right).
\eq
More relations can be found in \cite{Moch:1999eb,Gehrmann:2000zt}.

\end{appendix}

\end{document}